\documentclass[twocolumn,aps,superscriptaddress]{revtex4}

\usepackage{amssymb}
\usepackage{amsmath}
\usepackage{mathrsfs}
\usepackage{graphicx}
\usepackage{subfigure}
\usepackage[normalem]{ulem}
\usepackage[dvips]{color}

\begin{document}

\title{Properties of strange quark stars with isovector interactions}
\author{He Liu\footnote{liuhe@sinap.ac.cn}}
\affiliation{Shanghai Institute of Applied Physics, Chinese Academy of Sciences, Shanghai 201800, China}
\affiliation{University of Chinese Academy of Sciences, Beijing 100049, China}
\affiliation{Cyclotron Institute and Department of Physics and Astronomy, Texas A$\&$M University, College Station, TX 77843-3366, USA}
\date{\today}
\author{Jun Xu\footnote{corresponding author: xujun@zjlab.org.cn}}
\affiliation{Shanghai Advanced Research Institute, Chinese Academy of Sciences, Shanghai 201210, China}
\affiliation{Shanghai Institute of Applied Physics, Chinese Academy of Sciences, Shanghai 201800, China}
\author{Che Ming Ko\footnote{ko@comp.tamu.edu}}
\affiliation{Cyclotron Institute and Department of Physics and Astronomy, Texas A$\&$M University, College Station, TX 77843-3366, USA}
\date{\today}

\begin{abstract}
We study the properties of strange quark stars by employing a 3-flavor Nambu-Jona-Lasinio model with both scalar-isovector and vector-isovector interactions. Using the constraint on the vector-isoscalar interaction strength obtained from the elliptic flow splitting between particles and their antiparticles in relativistic heavy-ion collisions, we investigate the dependence of the properties of strange quark stars on the vector-isovector and the scalar-isovector interactions, and compare the results with the state-of-art astrophysical constraints on the compact star radius and mass as well as its tidal deformability from the GW170817 event. Results from our study reinforce the prospect of using both heavy-ion collisions and astrophysical observations to provide constraints on the isovector coupling strength in quark matter and thus the quark matter equation of state as well as the QCD phase structure at finite isospin chemical potentials.
\end{abstract}

\maketitle
Understanding the equation of state (EOS) of dense and strongly interacting matter is one of the main goals of nuclear physics. Studies based on heavy-ion collision experiments in terrestrial laboratories and compact star properties from astrophysical observations have already led to significant constraints on the EOS of dense nucleonic matter. However, our knowledge on the EOS of quark matter, which is related to the phase structure of the quantum chromodynamics (QCD), is still limited. Although the scalar part of the quark matter properties can be constrained by the lattice QCD calculations~\cite{Kar01,Kar02}, its vector part or especially the isovector part, which is relevant to the properties of quark matter at high net baryon densities and isospin asymmetries, remains poorly understood. On the other hand, a recent study using the transport approach based on a 3-flavor Nambu-Jona-Lasinio (NJL) model has shown, that the strength of the vector-isoscalar interaction can be extracted from the relative elliptic flow ($v_2$) difference between protons and antiprotons as well as between $K^+$ and $K^-$ in relativistic heavy-ion collisions, carried out in the beam-energy scan program at the relativistic heavy-ion collider~\cite{Jxu14}. Also, a strong vector-isovector interaction seems to be needed to reproduce the $v_2$ difference between $\pi^+$ and $\pi^-$ with the same NJL transport approach at the same collision energies~\cite{Liu19}. Both the vector and isovector interactions have been shown to have important impacts on the quark matter EOS and the QCD phase structure~\cite{Asa89,Fuk08,Bra12,Chu15,Liu16}. It is thus of great interest to investigate whether the quark interaction extracted from heavy-ion collisions is compatible with astrophysical observations or whether additional information can be extracted from the latter.

The mass and radius of compact stars, which are measurable through various methods~\cite{Mil16}, are the main astrophysical observables that can be used to extract information on the EOS of strongly interacting matter~\cite{Lat07}. The measurement of J1614-2230 and J0348+0432 a few years ago had led to a precise determination of $1.97 \pm 0.04 M_{\odot}$~\cite{Dem10} and $2.01 \pm 0.04 M_{\odot}$ for their respective masses~\cite{Ant13}, putting a strong constraint on the stiffness of the EOS of strongly interacting matter. For the radius of compact stars with the canonical mass $M=1.4M_\odot$, they have been constrained to the range of $10.62~{\rm km}<R_{1.4}<12.83~{\rm km}$ from analyzing the quiescent low-mass X-ray binaries~\cite{Lat14}. More recently, the gravitational wave event GW170817~\cite{Abb171} and its associated electromagnetic counterpart~\cite{Abb172} have provided additional constraints on the EOS of dense matter. The latest analysis of GW170817 by the LIGO+Virgo Collaboration has found with
a $90\%$ confidence that the tidal polarizability of the merging neutron stars is within the range $70<\Lambda_{1.4}<580$~\cite{Abb18}. In the present study, we will employ the above constraints on the compact star mass, radius, and tidal polarizability to extract information on the properties of dense matter.

Although the EOS of a pure nucleonic matter can generally be stiff enough to support a two-solar-mass neutron star, hyperon as well as quark degrees of freedom are expected to appear with the increasing baryon chemical potential, forming the so-called hybrid stars. It has been found that the high-mass constraint may be used to understand the properties of the hadron-quark phase transition as well as the EOS of the mixed phase in hybrid stars (see, e.g., Refs.~\cite{Sho03,Alf13,Bay18,Alv19,Han19}). Also, the radius of compact stars has been shown to be closely related to the isovector part of the EOS of dense matter~\cite{Li06}. Another possible configuration for compact stars is the strange quark star~\cite{Ito70,Hae86,Ter16} formed from the strange quark matter, i.e., $u$, $d$, and $s$ quarks as well as leptons under the charge neutrality and $\beta$-equilibrium conditions~\cite{Bom04,Sta07,Her11}. The properties of strange quark matter, which has been conjectured as the true ground state of QCD~\cite{Wit84,Far84} if its EOS is softer than the hadronic matter EOS, are governed by quark interactions, and they are also relevant to the QCD phase structure. In the present study, we investigate the properties of strange quark matter as well as strange quark stars based on the 3-flavor NJL model. With the vector-isoscalar coupling strength favored by the $v_2$ splitting between protons and antiprotons as well as between $K^+$ and $K^-$ in relativistic heavy-ion collisions~\cite{Jxu14}, and also considering our latest study on the $v_2$ difference between $\pi^+$ and $\pi^-$~\cite{Liu19}, we study in detail the properties of strange quark stars with vector-isovector and scalar-isovector interactions. Using data from astrophysical observations and heavy-ion experiments, we can obtain the constraints on the strengths of the quark vector and isovector interactions in quark matter.




The Lagrangian of the 3-flavor NJL model with isovector interactions can be expressed as~\footnote{The signs for vector interaction terms are set to be negative to be consistent with the relativistic mean-field model, different from those in Ref.~\cite{Liu16}.}
\begin{eqnarray}
\mathcal{L}_{\rm NJL} &=& \bar{q}(i\rlap{\slash}\partial-\hat{m})q
+\frac{G_S}{2}\sum_{a=0}^{8}[(\bar{q}\lambda_aq)^2+(\bar{q}i\gamma_5\lambda_aq)^2]
\notag\\
&-&\frac{G_V}{2}\sum_{a=0}^{8}[(\bar{q}\gamma_\mu\lambda_aq)^2+
(\bar{q}\gamma_5\gamma_\mu\lambda_aq)^2]
\notag\\
&-&K\{\det[\bar{q}(1+\gamma_5)q]+\det[\bar{q}(1-\gamma_5)q]\}
\notag\\
&+&G_{IS}\sum_{a=1}^{3}[(\bar{q}\lambda_aq)^2+(\bar{q}i\gamma_5\lambda_aq)^2]\notag\\
&-&G_{IV}\sum_{a=1}^{3}[(\bar{q}\gamma_\mu\lambda_aq)^2+(\bar{q}\gamma_5\gamma_\mu\lambda_aq)^2],
\end{eqnarray}
where $q = (u, d, s)^T$ and $\hat{m} = \text{diag}(m_u,m_d,m_s)$ are the quark fields and the current quark mass matrix for $u$, $d$, and $s$ quarks, respectively; $\lambda_a$ are the Gell-Mann matrices with $\lambda_0$ = $\sqrt{2/3}I$ in the 3-flavor space with the SU(3) symmetry; $G_S$ and $G_V$ are, respectively, the scalar-isoscalar and the vector-isoscalar coupling constant; and the $K$ term represents the six-point Kobayashi-Maskawa-t'Hooft interaction that breaks the axial $U(1)_A$ symmetry~\cite{Hoo76}. The additional $G_{IS}$ and $G_{IV}$ terms represent the scalar-isovector and the vector-isovector interactions, with $G_{IS}$ and $G_{IV}$ the corresponding coupling constants, respectively. Since the Gell-Mann matrices with $a = 1, 2, 3$ are identical to the Pauli matrices in $u$ and $d$ space, the isovector couplings break the SU(3) symmetry while keeping the isospin symmetry. In the present study, we employ the parameters $m_u = m_d = 3.6$ MeV, $m_s = 87$ MeV, $G_S\Lambda_c^2 = 3.6$, $K\Lambda_c^5 = 8.9$, and the cutoff value $\Lambda_c = 750$ MeV in the momentum integral given in Refs.~\cite{Bra12,Lut92}.

For the ease of discussions, we define the relative strength of the vector-isoscalar coupling, the scalar-isovector coupling, and the vector-isovector coupling respectively as $R_V=G_V/G_S$, $R_{IS}=G_{IS}/G_S$, and $R_{IV}=G_{IV}/G_S$. As is known, the position of the critical point for the chiral phase transition is sensitive to $R_V$~\cite{Asa89,Fuk08,Bra12}, which was later constrained within $0.5<R_V<1.1$ from the relative $v_2$ splitting between protons and antiprotons as well as between $K^+$ and $K^-$ in relativistic heavy-ion collisions~\cite{Jxu14}. $R_{IS}$ and $R_{IV}$ have also been shown to have dramatic effects on the phase diagram at larger isospin asymmetries as well as the quark matter symmetry energy~\cite{Liu16}. The vector and the isovector couplings are expected to affect the EOS and properties of quark matter at high baryon and isospin chemical potentials, respectively. We note that the effects of isoscalar interactions in the NJL model on the quark matter EOS and properties of strange quark stars have been investigated in Refs.~\cite{Han01,Wan19}. In this work, we will investigate the role of isovector interactions within the constraint of $0.5<R_V<1.1$.

In the mean-field approximation~\footnote{We choose the flavor-singlet state by considering only the $\lambda_0$ term in the vector-isoscalar interaction, to be consistent with Ref.~\cite{Jxu14} but different from that in Refs.~\cite{Chu15,Liu16}.}, the energy density $\varepsilon_Q$ of quark matter from the above NJL Lagrangian can be written as
\begin{eqnarray}\label{epsilon}
\varepsilon_Q&=&-2N_c\sum_{i=u,d,s}\int_0^{\Lambda_c}\frac{d^3p}{(2\pi)^3}
E_i(1-f_i-\bar{f}_i)
\notag\\
&-&\sum_{i=u,d,s}(\tilde{\mu}_i-\mu_i)\rho_i+G_S(\sigma_u^2+\sigma_d^2+\sigma_s^2)
\notag\\
&-&4K\sigma_u\sigma_d\sigma_s-\frac{2}{3}G_V(\rho_u+\rho_d+\rho_s)^2
\notag\\
&+&G_{IS}(\sigma_u-\sigma_d)^2-G_{IV}(\rho_u-\rho_d)^2-\varepsilon_0.
\end{eqnarray}
In the above, the factor $2N_c=6$ represents the spin and color degeneracy of the quark, $\varepsilon_0$ is introduced to ensure $\varepsilon_{Q}=0$ in vacuum, and $f_i$ and $\bar{f}_i$ are respectively the Fermi-Dirac distribution functions of quarks and antiquarks with flavor $i$. $\sigma_i = \langle q_i \bar{q_i}\rangle$ stands for the quark condensate related to the distribution function via
\begin{eqnarray}
\sigma_i&=&-2N_c\int_0^{\Lambda_c}\frac{d^3p}{(2\pi)^3}\frac{M_i}{E_i}(1-f_i-\bar{f}_i),
\end{eqnarray}
where $E_i=\sqrt{p^2 +{M_i}^2}$ is the single-quark energy, and the Dirac mass (or the constituent mass) $M_i$ of quark flavor $i$ is related to the quark condensate through the relation
\begin{eqnarray}\label{mass}
M_i &=&
m_i-2G_S\sigma_i+2K\sigma_j\sigma_k-2G_{IS}\tau_{3i}(\sigma_u-\sigma_d),
\label{mi}
\end{eqnarray}
with ($i$, $j$, $k$) being any permutation of ($u$, $d$, $s$) and $\tau_{3i}$ being the isospin quantum number of quarks, i.e., $\tau_{3u} = 1$, $\tau_{3d} = -1$, and $\tau_{3s} = 0$. The net number density $\rho_i$ of quark flavor $i$ can be calculated from the quark and antiquark distribution functions via
\begin{eqnarray}
\rho_i=2N_c\int^{\Lambda_c}_0(f_i-\bar{f_i})\frac{d^3p}{(2\pi)^3}.
\end{eqnarray}
The relation between the effective chemical potential $\tilde{\mu}_i$ and the real chemical potential $\mu_i$ depends on the vector interactions, i.e.,
\begin{eqnarray}\label{mui}
\tilde{\mu}_i&=&\mu_i-\frac{4}{3}G_V(\rho_u+\rho_d+\rho_s)-2G_{IV}\tau_{3i}(\rho_u-\rho_d).
\end{eqnarray}
At zero temperature relevant to the present study, the quark distribution reduces to the step function $f_i=\Theta(\tilde{\mu}_i-E_i)$, and antiquarks disappear, i.e., $\bar{f}_i=0$. The above equations can be solved self-consistently, and from which the pressure of quark matter at zero temperature can be obtained from
\begin{equation}
P_Q=\sum_{i=u, d, s}\mu_i\rho_i-\varepsilon_Q.
\label{eq2}
\end{equation}


The strange quark star is composed of a mixture of quarks ($u$, $d$, and $s$) and leptons ($e$ and $\mu$) under the charge neutrality condition
\begin{equation}
\frac{2}{3}\rho_u-\frac{1}{3}(\rho_d+\rho_s)-\rho_e-\rho_\mu=0,
\end{equation}
and the $\beta$-equilibrium condition
\begin{eqnarray}
\mu_s&=&\mu_d=\mu_u+\mu_e,
\\
\mu_\mu&=&\mu_e.
\end{eqnarray}
In terms of the electron mass $m_e=0.511$ MeV and the muon mass $m_\mu=106$ MeV, the lepton contributions to the energy density and the pressure are
\begin{eqnarray}
\varepsilon_L&=&\sum_{i=e, \mu}\frac{1}{\pi^2}\int_0^{p_F^i}\sqrt{p^2+m_i^2}p^2dp,
\\
P_L&=&\sum_{i=e, \mu}\mu_i\rho_i-\varepsilon_L,
\end{eqnarray}
where $p_F^i=(3\pi\rho_i/2)^\frac{1}{3}$ is the lepton Fermi momentum. The total energy density and pressure including the contributions from both quarks and leptons are
\begin{eqnarray}
\varepsilon&=&\varepsilon_Q+\varepsilon_L,
\\
P&=&P_Q+P_L,
\end{eqnarray}
where $\varepsilon_Q$ and $P_Q$ are from the NJL model given above.

Using the pressure $P$ and the energy density $\varepsilon$, the mass-radius relation of compact stars can be obtained by solving the Tolman-Oppenheimer-Volkoff equation, i.e.,
\begin{eqnarray}
\frac{dP(r)}{dr}&=&-\frac{M(r)[\varepsilon(r)+P(r)]}{r^2}\left[1+\frac{4 \pi P(r)r^3}{M(r)}\right]\notag\\
&\times&\left[1-\frac{2M(r)}{r}\right]^{-1},
\end{eqnarray}
where $M(r)$ is the gravitational mass inside the radius $r$ of the compact star and can be obtained from the integral of the following equation
\begin{eqnarray}
\frac{dM(r)}{dr}&=&4 \pi r^2 \varepsilon(r).
\end{eqnarray}

The gravitational waves emitted from the merge of two compact stars serve as another probe to the EOS of dense matter~\cite{Hin08,Rea09}. The tidal deformability $\Lambda$ of compact stars during their merge is related to the love number $k_2$ through the relation $\Lambda = \frac{2}{3}k_2\beta^{-5}$, with the latter given by~\cite{Hin08,Pos10}
\begin{eqnarray}
k_2 &=&\frac{8}{5}(1-2\beta)^2[2-y_R+2\beta(y_R-1)]
\notag\\
&\times& \{2\beta[6-3y_R+3\beta(5y_R-8)]
\notag\\
&+& 4\beta^3[13-11y_R+\beta(3y_R-2)+2\beta^2(1+y_R)]
\notag\\
&+&3(1-2\beta)^2[2-y_R+2\beta(y_R-1)]\text{ln}(1-2\beta)\}^{-1}.\notag\\
\end{eqnarray}
In the above, $\beta \equiv M/R $ is the compactness of the compact star, and $y_R \equiv y(R) $ is the solution at the star surface to the first-order differential equation
\begin{eqnarray}
r \frac{dy(r)}{dr}+y(r)^2+y(r)F(r)+r^2Q(r)=0,
\end{eqnarray}
with
\begin{eqnarray}
F(r) &=& \frac{r-4\pi r^3[\varepsilon(r)-P(r)]}{r-2M(r)},
\notag\\
Q(r)&=&\frac{4\pi r\left[5\varepsilon(r)+9P(r)+\frac{\varepsilon(r)+P(r)}{\partial P(r)/\partial \varepsilon(r)}-\frac{6}{4\pi r^2}\right]}{r-2M(r)}
\notag\\
&-&4\left[\frac{M(r)+4\pi r^3P(r)}{r^2(1-2M(r)/r)}\right]^2.
\end{eqnarray}
For a given central density $\rho_c$ and using the boundary conditions $y(0) = 2$, $M(0)=0$, $P(0)=P_c$, and $\varepsilon(0)=\varepsilon_c$, the above differential equations can be solved, and the mass $M$, radius $R$, and the tidal deformability $\Lambda$ can be obtained. By changing the value of $\rho_c$, one gets series of strange quark stars with relations among $M$, $R$, and $\Lambda$ based on a particular strange quark matter EOS.


\begin{figure}[tbh]
\includegraphics[scale=0.3]{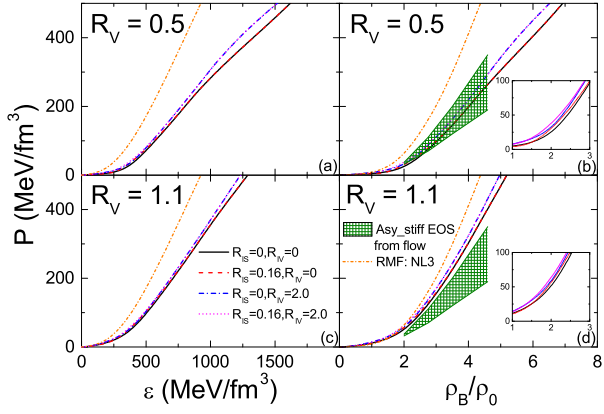}
\caption{(color online) The equation of state, i.e., pressure as a function of the energy density (left) and the reduced baryon density (right) for strange quark matter with different coupling strengths of the vector-isovector interaction $R_{IV}$ and the scalar-isovector interaction $R_{IS}$, for a given vector-isoscalar interaction coupling strength $R_V=0.5$ (upper) and 1.1 (lower). Here $\rho_B=(\rho_u+\rho_d+\rho_s)/3$ is the baryon density, and $\rho_0=0.16$ fm$^3$ is the saturation density of normal nuclear matter. The Asy\_stiff EOS band~\cite{Dan02} for neutron matter from the flow data as well as the neutron star matter EOS from the NL3 parameterization~\cite{Lal97} of the relativistic mean-field model (RMF: NL3) are also shown for comparison.}
\label{fig1}
\end{figure}

We first discuss the impact of the isovector interactions on the EOS of strange quark matter from the 3-flavor NJL model for $0.5<R_V<1.1$. As shown in Fig.~\ref{fig1}, the EOS is more sensitive to the strength of the vector-isoscalar interaction through the $\frac{2}{3}G_V(\rho_u+\rho_d+\rho_s)^2$ term in Eq.~\eqref{epsilon}, with a larger $R_V$ leading to a stiffer EOS for the strange quark matter, consistent with that observed in Ref.~\cite{Han01}. The vector-isovector interaction characterized by the reduced coupling constant $R_{IV}$ slightly stiffens the EOS at higher baryon densities, since its contribution is determined by the $G_{IV}(\rho_u-\rho_d)^2$ term in Eq.~\eqref{epsilon}. For the scalar-isovector interaction characterized by the reduced coupling constant $R_{IS}$, it affects the EOS only around $2\rho_0$ as shown in the insets, but has negligible effects at higher baryon densities. The Asy\_stiff EOS band~\cite{Dan02} for neutron matter from the flow data is expected to be similar to that for the corresponding neutron star matter, providing a standard EOS uncertainty range for the hadronic matter. The existence of the strange quark matter requires its EOS be below the upper boundary of the hadronic matter EOS, and this gives some constraints on the NJL coupling constants to be discussed later. As for the EOS of the neutron star matter from the NL3 parameterization~\cite{Lal97} of the relativistic mean-field model, which is extensively used in astrophysics studies, it is stiffer than any EOSs from different NJL coupling constants used in the present study.

\begin{figure}[tbh]
\includegraphics[scale=0.3]{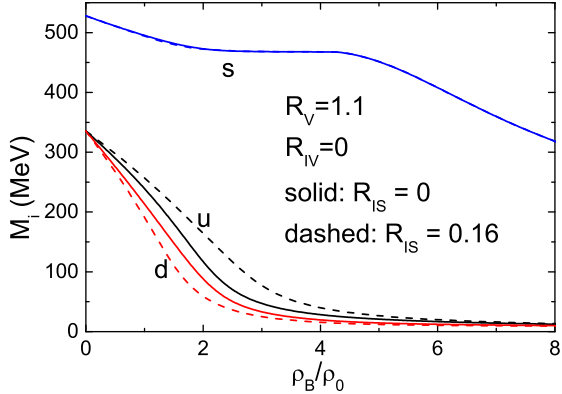}
\caption{(color online) Constituent masses of $u$, $d$, and $s$ quarks as functions of the reduced baryon density in strange quark matter for different coupling strengths of the scalar-isovector interaction $R_{IS}=0$ and 0.16. }
\label{fig2}
\end{figure}

To understand the effect of the scalar-isovector interaction, we display in Fig.~\ref{fig2} how the constituent mass of various quarks in strange quark matter changes with its reduced baryon density. The decrease of constituent mass with increasing baryon density is observed for all quark species as a result of chiral symmetry restoration at finite baryon densities. In the isospin asymmetric strange quark matter with different $u$ and $d$ quark densities, the mass splitting between $u$ and $d$ quarks is observed even without the scalar-isovector interaction as a result of the $K$ term in Eq.~\eqref{mass}. In the presence of the scalar-isovector interaction, the mass splitting between $u$ and $d$ quarks is enhanced due to the $G_{IS}$ term in Eq.~\eqref{mass}. Further increasing the strength of the scalar-isovector interaction may lead to too large a mass splitting and thus an unreasonable negative $d$ quark mass. We note that the scalar-isovector interaction affects the EOS mostly around $2\rho_0$ through the mass splitting of $u$ and $d$ quarks, as shown in Fig.~\ref{fig1}.

\begin{figure}[tbh]
\includegraphics[scale=0.3]{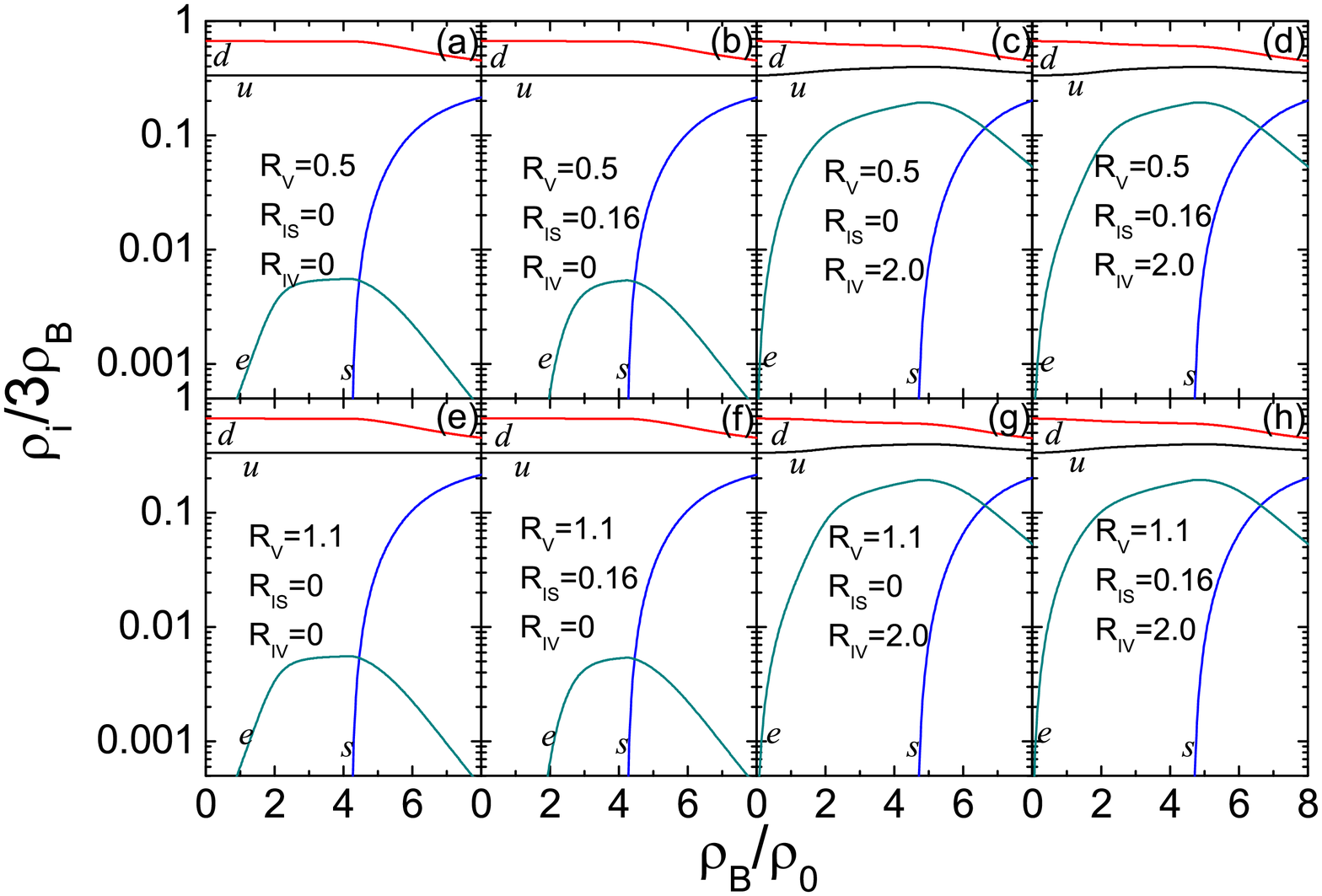}
\caption{(color online) Relative fractions of $u$, $d$, $s$, and $e$ components as functions of the reduced baryon density in strange quark matter for different coupling strengths of the vector-isoscalar, scalar-isovector, and vector-isovector interactions.}
\label{fig3}
\end{figure}

To better understand the effects of isovector interactions on the EOS of strange quark matter, we show in Fig.~\ref{fig3} the relative fraction of various quarks and leptons for different coupling strengths of these interactions.  The decrease of the $d$ quark and electron fractions when $s$ quarks appear is observed in all cases. With the mean-field approximation for only the flavor-singlet state, the vector-isoscalar interaction has negligible effects on the relative fractions of various particles. As to the vector-isovector interaction, it enhances significantly the electron fraction and reduces the electron threshold density, and at the same time reduces the difference between the $u$ and $d$ quark fractions. The latter is understandable since the vector-isovector interaction tends to reduce the isospin asymmetry in strange quark matter and thus the energy of the system. For the scalar-isovector interaction, it affects only slightly the electron threshold density. Due to the small isospin/charge chemical potential in the strange quark matter, no muons are present at all densities.

\begin{figure}[tbh]
\includegraphics[scale=0.3]{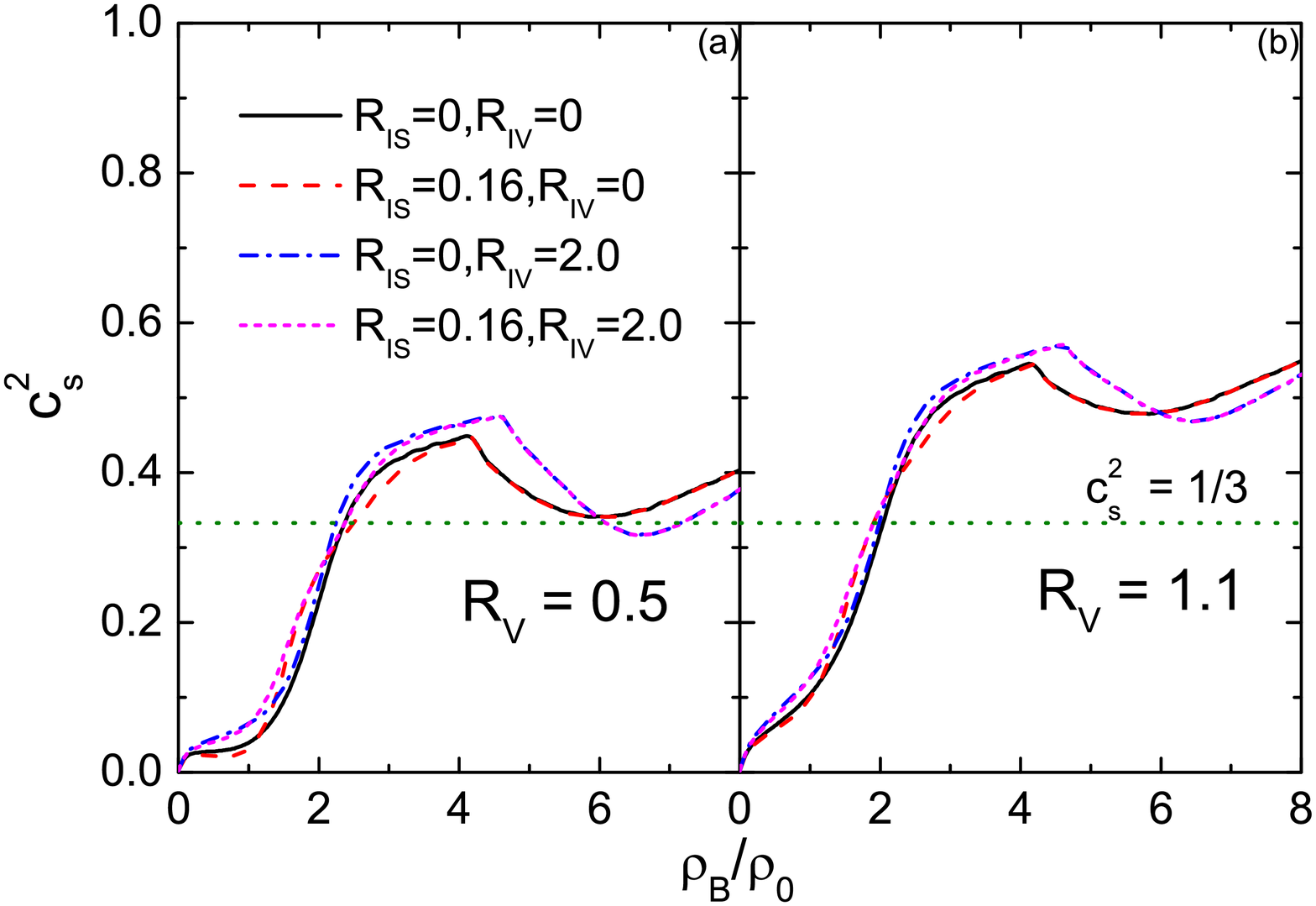}
\caption{(color online) The square of the sound velocity as a function of the reduced baryon density in strange quark matter for different coupling strengths of the vector-isovector interaction $R_{IV}$ and the scalar-isovector interaction $R_{IS}$, for a given vector-isoscalar interaction coupling strength $R_V=0.5$ (left) and 1.1 (right). }
\label{fig4}
\end{figure}

The sound velocity $c_s$, which can be calculated from $c_s^2=\partial P/\partial\epsilon$, is another property of the strange quark matter, and it can be used to check if the underlying EOS satisfies the causality condition. As shown in Fig.~\ref{fig4}, a stronger vector-isoscalar interaction leads to a quicker increase of the sound velocity with increasing baryon density as expected. The peak of the sound velocity occurs at a baryon density where $s$ quarks appear and thus soften the EOS as a result of more degrees of freedom. As shown in Fig.~\ref{fig3}, the threshold density for the appearance of $s$ quarks, which corresponds to the peak of the sound velocity, is sensitive to the strength of the vector-isovector interaction. The sound velocity further increases at even higher densities, and this is due to the disappearance of electrons and the saturation of the $s$ quark fraction, as can also be seen from Fig.~\ref{fig3}. Similar to the effect on the EOS, the scalar-isovector interaction affects the sound velocity only around $2\rho_0$. Also shown in Fig.~\ref{fig3} is the sound velocity $c_s^2=1/3$ in the conformal limit corresponding to free massless fermions, and it is seen that our results with a strong repulsive vector-isoscalar interaction are larger than this limit at higher densities, indicating that the corresponding EOS is stiffer than that of massless fermions. We note that for all densities considered here the causality condition is safely satisfied.

\begin{figure}[tbh]
\includegraphics[scale=0.3]{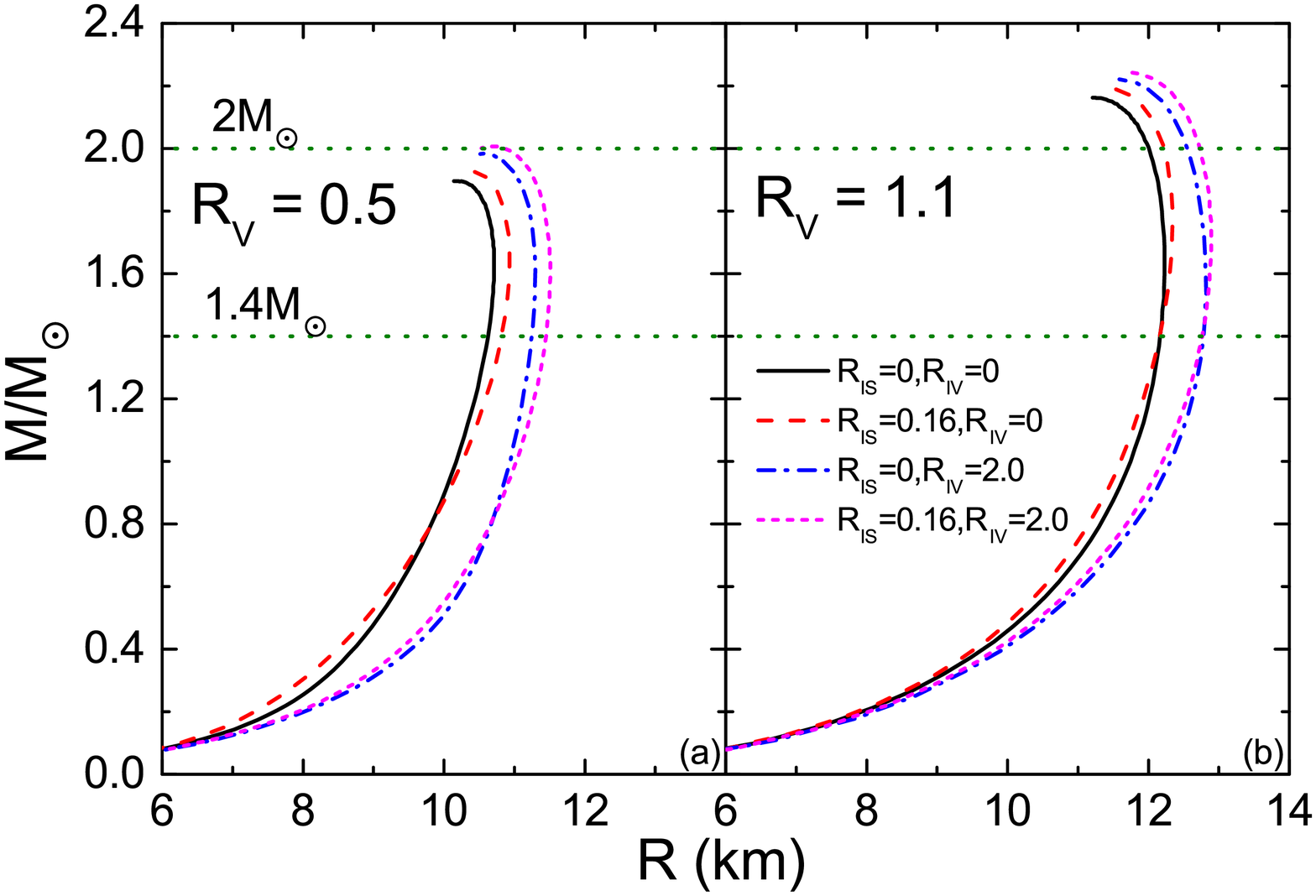}
\caption{(color online) Mass-radius relation for strange quark stars with different coupling strengths for the vector-isovector interaction $R_{IV}$ and the scalar-isovector interaction $R_{IS}$, for a given vector-isoscalar interaction coupling strength $R_V=0.5$ (left) and 1.1 (right).}
\label{fig5}
\end{figure}

\begin{figure}[tbh]
\includegraphics[scale=0.3]{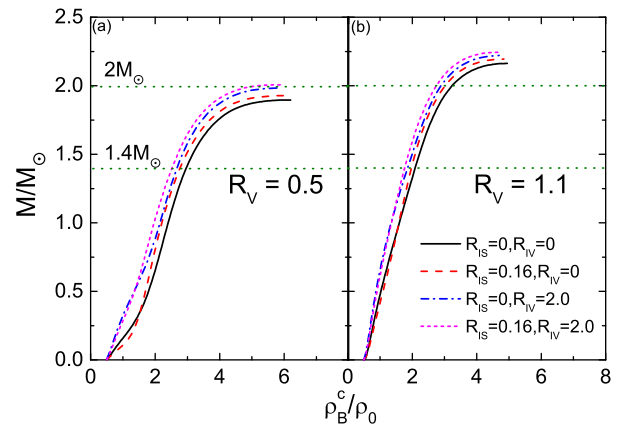}
\caption{(color online) Relation between the mass and the reduced central baryon density of strange quark stars with different coupling strengths of the isovector interaction $R_{IV}$ and the scalar-isovector interaction $R_{IS}$, for a given vector-isoscalar interaction coupling strength $R_V=0.5$ (left) and 1.1 (right). }
\label{fig6}
\end{figure}

The relations between the mass and radius of a strange quark star as well as between its mass and central baryon density from different scenarios are compared in Figs.~\ref{fig5} and \ref{fig6}, respectively. Without isovector couplings, the maximum mass of strange quark stars can reach about $1.8M_{\odot}$ for $R_V=0.5$ and $2.2M_{\odot}$ for $R_V=1.1$. Including the vector-isovector interaction increases both the maximum mass and the radius of the strange quark star, while including the scalar-isovector interaction increases slightly the maximum mass but has different effects on the radius for stars with higher or lower masses. These can be understood from the EOSs for different scenarios as shown in Fig.~\ref{fig1}, and the sensitivity of the compact star radius to the isovector part of the EOS is consistent with that observed in Ref.~\cite{Li06}. After the mass of compact stars reaches the maximum value, further increasing the central density will reduce both the maximum mass and the radius, making the compact star unstable. Results corresponding to such unphysical regions are not plotted in these figures, and our study indicates that the maximum baryon density for strange quark stars is less than $6\rho_0$.

\begin{figure}[tbh]
\includegraphics[scale=0.3]{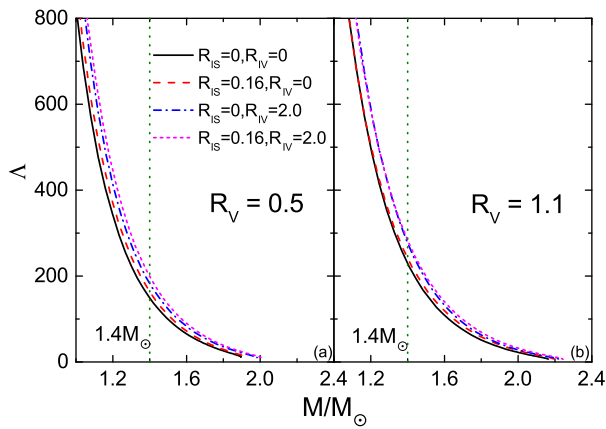}
\caption{(color online) Relation between the tidal deformability and the mass of strange quark stars with different coupling strengths of the vector-isovector interaction $R_{IV}$ and the scalar-isovector interaction $R_{IS}$, for a given vector-isoscalar interaction coupling strength $R_V=0.5$ (left) and 1.1 (right). }
\label{fig7}
\end{figure}

The tidal deformability as a function of strange quark star mass for different scenarios is displayed in Fig.~\ref{fig7}. The general behavior that the deformability decreases with increasing mass is observed, as long as the strange quark star is stable with increasing central density. For a given mass, the deformability is larger for a larger radius, as expected. Thus, the value $R_V=1.1$ generally gives a larger deformability than the value $R_V=0.5$ for a canonical star mass of $M=1.4M_\odot$. Including the isovector interactions also increases slightly the deformability, consistent with the slightly larger radius observed in Fig.~\ref{fig5} when isovector interactions are included. For such a canonical star mass, the tidal deformabilities from all scenarios are within the uncertainty range $70<\Lambda_{1.4}<580$ extracted from the GW170817 event~\cite{Abb18}. The large uncertainty in the tidal deformability thus cannot give additional constraints on quark interactions. If the second gravitational wave from a more recently observed compact binary coalescence with a total mass $\sim 3.4 M_\odot$~\cite{LIGO20} is measured, or the new results from the Neutron Star Interior Composition Explorer (NICER) are available, they may be able to put a more stringent constraint on the quark matter EOS.

\begin{figure}[tbh]
\includegraphics[scale=0.3]{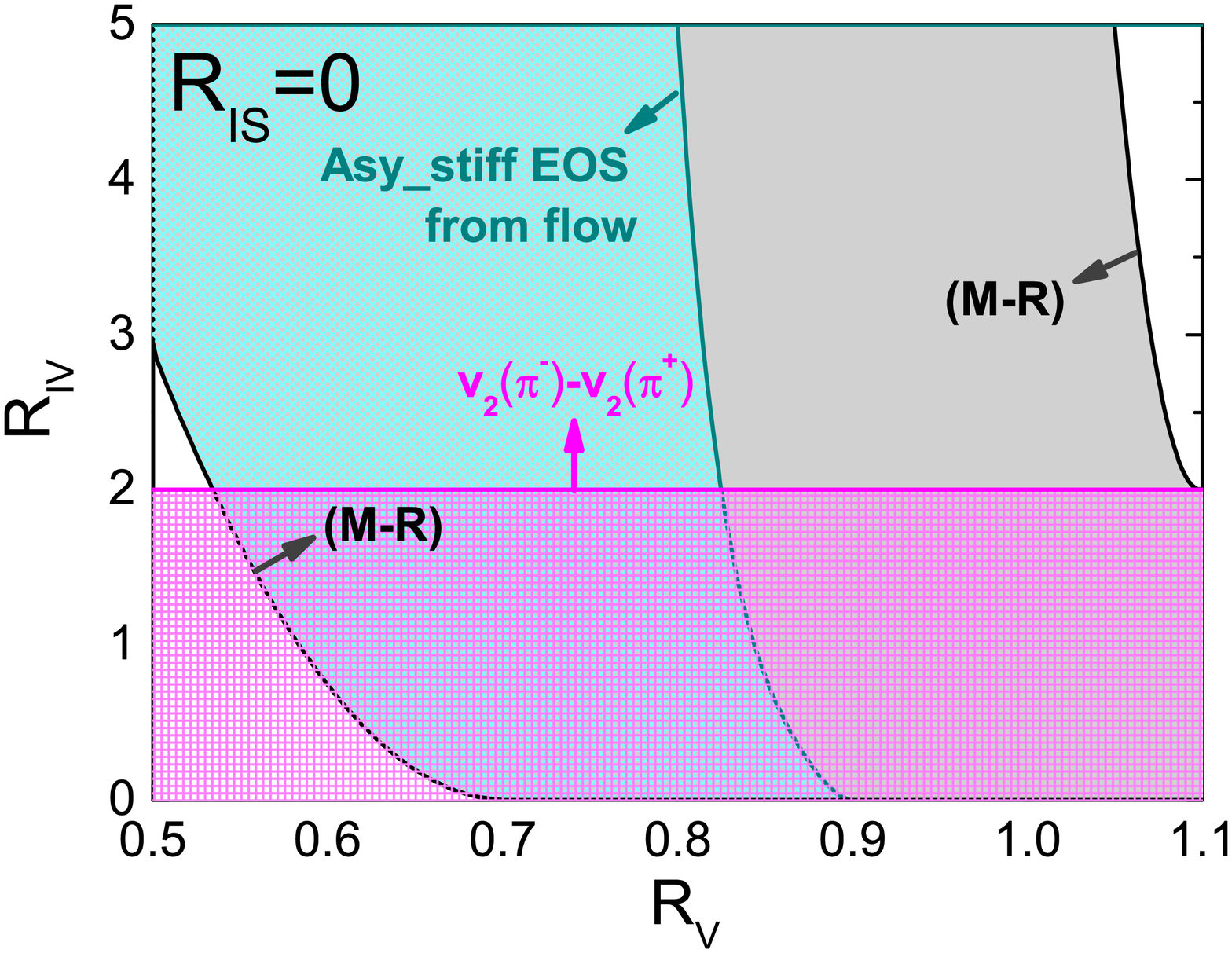}\\
\includegraphics[scale=0.3]{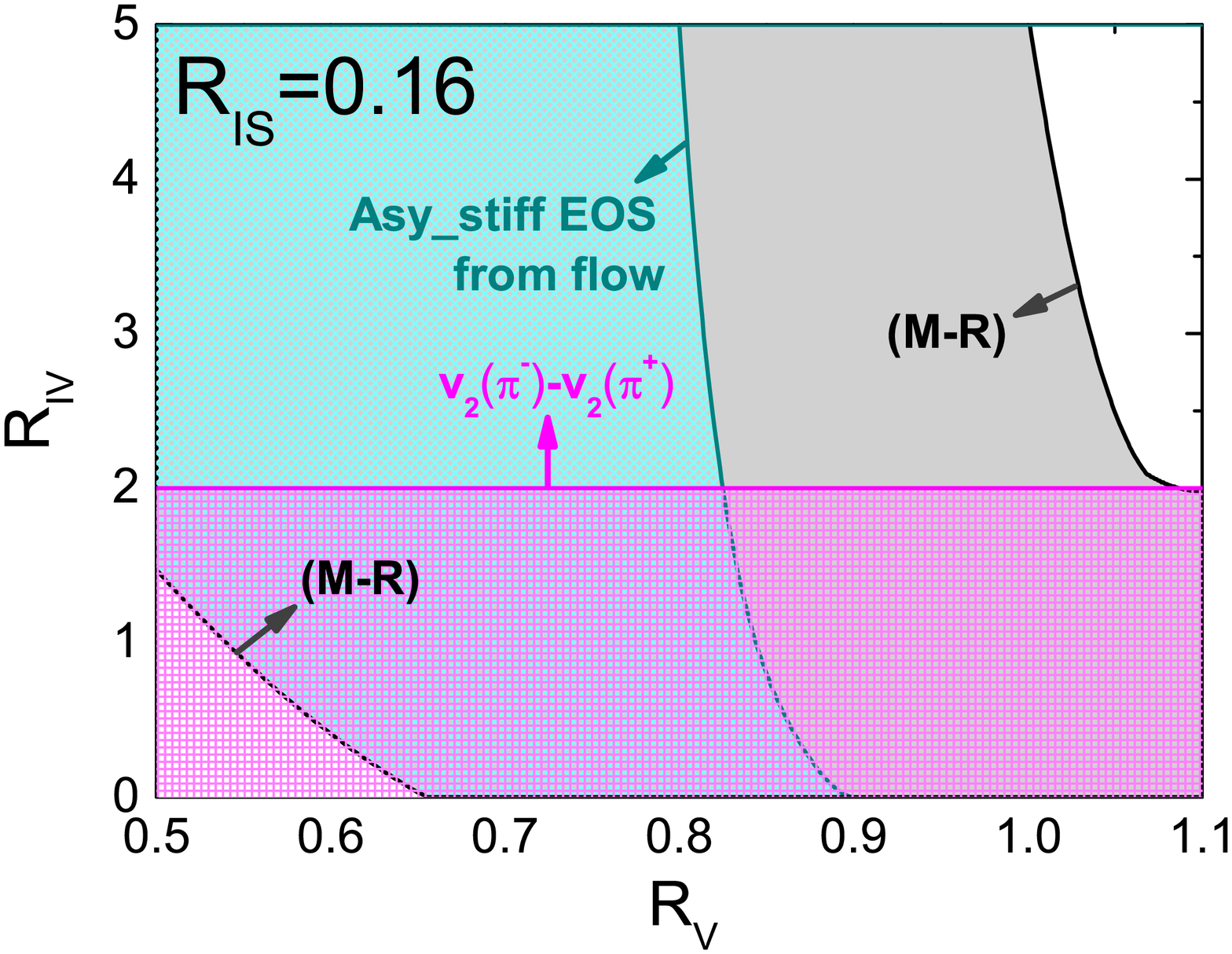}
\caption{(color online) Constraints on the strengths of the vector-isoscalar interaction $R_V$ and the vector-isovector interaction $R_{IV}$ from comparing with the hadronic matter EOS, the $v_2$ splitting results in relativistic heavy-ion collisions, and the mass-radius relation for strange quark stars. The upper panel is the case without the scalar-isovector interaction $R_{IS}=0$, while the lower panel is for $R_{IS}=0.16$. }
\label{fig8}
\end{figure}

We summarize the constraints on the coupling strengths of the vector-isoscalar and the vector-isovector interactions from heavy-ion collisions and astrophysical observations in Fig.~\ref{fig8}. Considering the uncertainties of the scalar-isovector interaction, constraints for both $R_{IS}=0$ and $R_{IS}=0.16$ are plotted. As shown in Fig.~\ref{fig1}, the stability condition for the strange quark matter requires its EOS be below the upper boundary of the hadronic matter EOS represented by the Asy\_stiff EOS band from the flow analysis, and this automatically gives some constraints on the NJL coupling constants as shown in Fig.~\ref{fig8}. The range $0.5<R_V<1.1$ is from the $v_2$ splitting between protons and antiprotons as well as between $K^+$ and $K^-$ in relativistic heavy-ion collisions~\cite{Jxu14}, while our more recent study on the $v_2$ splitting between $\pi^+$ and $\pi^-$ favors $R_{IV}>2$~\cite{Liu19} and is rather insensitive to $R_V$ and $R_{IS}$. As shown in Fig.~\ref{fig5}, the vector-isovector interaction is needed to give the maximum mass of strange quark star as large as $2M_{\odot}$ for smaller $R_V$, while it may lead to too large a radius for a canonical compact star compared to the constraint $10.62~{\rm km}<R_{1.4}<12.83~{\rm km}$ for larger $R_V$. Given that the stability condition for the strange quark matter is satisfied, the mass-radius relation of strange quark stars together with the $v_2$ splitting results in relativistic heavy-ion collisions thus gives an allowed area for $(R_V,R_{IV})$, as shown in Fig.~\ref{fig8}, for different values of $R_{IS}$.


In conclusion, we have studied the properties of strange quark stars by employing the 3-flavor Nambu-Jona-Lasinio model with isovector interactions included and the vector-isoscalar interaction constrained by the elliptic flow splitting in relativistic heavy-ion collisions. Effects of the isovector interactions on the equation of state, the $u$ and $d$ quark mass splitting, the relative multiplicity fraction for different particle species, and the sound velocity in strange quark matter are discussed. Although the resulting tidal deformabilities are all consistent with the uncertainty range extracted from the GW170817 event, the state-of-art constraints on the compact mass and radius limit the coupling strengths of the isovector interactions.  An allowed region for the strengths of the vector-isoscalar, vector-isovector, and scalar-isovector interaction is obtained from both elliptic flow splitting between particles and antiparticles in relativistic heavy-ion collisions and astrophysical observations, given that the stability condition of strange quark matter is satisfied. This may help to better understand the quark matter equation of state as well as the QCD phase structure at finite isospin chemical potentials.

J.X. acknowledges support from the National Natural Science Foundation of China under Grant No. 11922514. C.M.K. acknowledges support from the US Department of Energy under Contract No. DE-SC0015266 and the Welch Foundation under Grant No. A-1358.

\end{document}